\documentclass{pasj00}
\usepackage{bm}

\usepackage[dvips]{color}

\newcommand{\argmin}[1]{\mathop{\rm arg~min}_{#1}\limits}

\begin{document}
\SetRunningHead{Uemura, et al.}{Variable selection for SN Ia supernovae}

\title{Variable Selection for Modeling the Absolute Magnitude at
  Maximum of Type~Ia Supernovae}

\author{Makoto \textsc{Uemura}, Koji, S. \textsc{Kawabata}} %
\affil{Hiroshima Astrophysical Science Center, Hiroshima University,
  Kagamiyama 1-3-1, Higashi-Hiroshima, 739-8526, Japan}
\affil{Core of Research for the Energetic Universe, Hiroshima
  University, Higashi-Hiroshima, Hiroshima 739-8526, Japan}
\email{uemuram@hiroshima-u.ac.jp}

\author{Shiro \textsc{Ikeda}}
\affil{The Institute of Statistical Mathematics and CREST, JST,
  Tachikawa, Tokyo, 190-8562, Japan} 

\and
\author{Keiichi \textsc{Maeda}}
\affil{Department of Astronomy, Kyoto University,
  Kitashirakawa-Oiwake-cho Sakyo-ku, Kyoto 606-8502, Japan}
\affil{Kavli Institute for the Physics and Mathematics of the Universe
  (WPI), University of Tokyo, 5-1-5 Kashiwanoha, Kashiwa, Chiba
  277-8583, Japan} 


\KeyWords{supernovae: general} 

\maketitle

\begin{abstract}
We discuss what is an appropriate set of explanatory variables in
order to predict the absolute magnitude at the maximum of Type~Ia
supernovae. In order to have a good prediction, the error for future
data, which is called the ``generalization error,'' should be
small. We use cross-validation in order to control the generalization
error and LASSO-type estimator in order to choose the set of
variables. This approach can be used even in the case that the number
of samples is smaller than the number of candidate variables. We
studied the Berkeley supernova database with our approach. Candidates
of the explanatory variables include normalized spectral data,
variables about lines, and previously proposed flux-ratios, as well
as the color and light-curve widths. As a result, we confirmed the
past understanding about Type~Ia supernova: i)~The absolute magnitude
at maximum depends on the color and light-curve width. ii)~The
light-curve width depends on the strength of Si\,\textsc{ii}. 
Recent studies have suggested to add more variables in order to
explain the absolute magnitude. However, our analysis does not support
to add any other variables in order to have a better generalization
error. 
\end{abstract}

\section{Introduction}

Type Ia supernovae (SNe~Ia) have been used as ``standard candles'' to
estimate the distance to galaxies in cosmology. \citet{phi93law} found
a significant correlation between their absolute magnitude at maximum,
$M$, and decay rate, and proposed that a better distance indicator can
be obtained by calibrating it. As well as the decay rate, the observed
color also exhibits a clear correlation with $M$. This is mainly due to
the interstellar extinction in both their host and our galaxies, while
it is proposed that there is a variation in the intrinsic color of 
SNe~Ia at maximum (\cite{con07col,fol11vel}). In addition to these two,
a number of variables have been proposed as explanatory variables of
$M$. They are, for example, the equivalent widths, velocities, or
depths of absorption lines, or their ratios (for a review, see
\cite{bsnip3}). 

The search for a good set of variables, in other words, the ``model,'' 
have recently been intensified including arbitrary ratios of the
fluxes in spectra. Using the 58 objects observed by Nearby Supernova
Factory, \citet{bai09frat} report that the model with a single ratio
of the flux at 642~nm to that at 443~nm, hereafter
$\mathcal{R}(642\,{\rm nm}/443\,{\rm nm})$, has a smaller residual of
$M$ than the classical model with the color and decay rate (, or
light-curve width). Using 26 objects observed by the CfA Supernova
Program, \citet{blo11frat} confirm the conclusion in \citet{bai09frat}
with a slightly different ratio, $\mathcal{R}(6630\,{\rm \AA}/ 
4400\,{\rm \AA})$, although the improvement of the model has low
significance. In addition, they propose another model with the color
and the color-corrected flux ratio, $\mathcal{R}^c(4610\,{\rm
  \AA}/4260\,{\rm \AA})$ at $t=-2.5\,{\rm d}$ from maximum
light. \citet{bsnip3}, using 62 object observed by the Berkeley
Supernova~Ia Program, report that the best set of variables is the
light-curve width, color, and $\mathcal{R}^c(3780\,{\rm
  \AA}/4580\,{\rm \AA})$. On the other hand, their analysis did not
confirm the results in \citet{bai09frat} and
\citet{blo11frat}. Thus, the resulting models of each work are
not completely consistent, and the model for the prediction of $M$
has not been established.

In previous studies, a linear regression model of $M$ has been assumed: 
\begin{eqnarray}
M_B \simeq M_{B,0} + \beta_1 x_1 + \beta_2 x_2 + \cdots + \beta_L x_L,
\end{eqnarray}
where $M_B$ is the absolute magnitude in the $B$-band, which has been 
conventionally used in past studies. $M_{B,0}$ is a constant. 
The vector, $\bm{x}=(x_1, x_2, \cdots, x_L)^T$ is a set of explanatory
variables of $M_B$. The elements in $\bm{x}$ are, for example, the
color, decay rate (, or light-curve width), and variables about the
lines. $\bm{\beta}=(\beta_1, \beta_2,\cdots, \beta_L)^T$ is the vector
of their coefficients. Suppose that $N$ samples of SNe~Ia are
available, and the observations are summarized as $\bm{y}\simeq
X\bm{\beta}$, where $\bm{y}=(M_{B1},M_{B2},\cdots,M_{BN})^T$ and
$X=(\bm{x}_1,\cdots,\bm{x}_N)^T$. The goal of the study is to find an
appropriate set of variables in $\bm{x}$ for the prediction of $\bm{y}$.
We prefer the model to have a small generalization error for the
prediction of $\bm{y}$. 

If $N\geq L$, it is possible to estimate the values of all elements in
$\bm{\beta}$ with the least-square method. However, the risk of
over-fitting increases as $N/L$ becomes smaller. Furthermore, the
least-square method cannot determine a unique model when $N<L$. Such a
situation can appear when arbitrary flux ratios in spectra are
included into $X$. Hence, previous studies included only one or two
flux ratios in a model, and search for the best set of the variables 
for the observations. 

Finding an appropriate set of variables to describe $M_B$ of SNe~Ia is
a variable selection problem, which has been studied in the field of
statistics and machine learning. In this paper, we report a result of
variable selection approach applied for $M_B$. We controlled the
generalization error with a regularization term, whose size is chosen
via cross-validation, and a subset of the variables are selected from
$L$ components by Least Absolute Shrinkage and Selection Operator, or
the so-called, LASSO method (\cite{LASSO}). This method can find the
unique solution even in the case of $N<L$. In section~2, we describe
the method. In section~3, we report on the results of our
experiments. We apply our method to the data provided by the Berkeley
supernova database. In section~4, we discuss the implication of our
results, and summarize our findings. 

\section{Method}
\subsection{LASSO-type estimation}

Here, we consider a linear regression model, $\bm{y}=X\bm{\beta}+\bm{e}$, 
where $X$ is a given real $N\times L$ matrix and $\bm{e}$ is a Gaussian
noise with $E[\bm{e}]=\bm{0}$ and $E[\bm{e}\bm{e}']=\sigma^2I_N$. Our
goal is to find an appropriate set of variables from $L$ variables and
$N$ samples and compute the corresponding coefficients of
$\bm{\beta}$. For this sort of estimation problems, \citet{LASSO}
proposed a method, Least Absolute Shrinkage and Selection Operator, or
the so-called, LASSO, for selecting the best set of explanatory variables.
LASSO provides a solution $\bm{\hat{\beta}}$ by minimizing the
following function which includes the $\ell 1$-norm of $\bm{\beta}$ as
a regularization term
\begin{eqnarray}
\bm{\hat{\beta}}_\lambda = \argmin{\bm{\beta}} \left\{ \| \bm{y}-X\bm{\beta} \|^2_2 + \lambda \|
  \bm{\beta} \|_1 \right\},
\end{eqnarray}
where $\|\bm{\beta}\|_1$ is the $\ell 1$-norm, defined as 
$\|\bm{\beta}\|_1=\sum_i |\beta_i|$, and $\lambda$ is a tunable constant. 
The estimate $\bm{\hat{\beta}}$ includes $0$ components, that is,
variables selection is realized with LASSO-type estimation. The number
of $0$ components increases as $\lambda$ becomes larger. 

We apply the LASSO-type estimation in order to select an appropriate
model to predict $M_B$ of SNe~Ia. The data, $\bm{y}$, is $M_B$ and
each column of $X$ corresponds to an observed variables, such as, the
color, light-curve width, and variables about spectra. Recent 
projects have provided high-quality and uniform samples of SNe~Ia in
both photometric and spectroscopic data. The number of available
samples, $N$, is now $\sim 100$. The number of candidate explanatory
variables can be $>10^4$ if arbitrary flux ratios are
included. However, we can expect that the number of effective
variables is small. In other words, our interest focuses on a model in
which $M_B$ is explained not with $\sim 10^4$, but with only a few
variables of $\bm{x}$. Exhaustive search for every subset of candidate
variables is not tractable, and the LASSO-type estimation gives us a
data-driven approach to select the best subset of variables for the
data-set. 
 
\subsection{Cross-validation}

The cost function for the estimation expressed in equation~(2)
contains a tunable parameter, $\lambda$. This parameter controls the
weight of the regularization term, which has an influence on the
generalization error. We choose the best $\lambda$ by the
cross-validation method. In the $K$-fold cross-validation, the
data is divided into $K$ roughly equal sub-samples, $\bm{y}_k$
($k=1,2,\cdots,K$). For each $k$, the training data is defined as all
the $K-1$ sub-samples except for the validation data, $\bm{y}_k$. The
optimization of the model to the training data gives
$\hat{\bm{\beta}}_{k,\lambda}$ at a certain $\lambda$. The
generalization error of the model is evaluated with the mean of
weighted mean square errors (wMSE; $E(\lambda)$) of the $K$
sub-samples; 
\begin{eqnarray}
E(\lambda) &=& \frac{1}{K} \sum_{k=1}^K E_k(\lambda)\\
E_k(\lambda) &=& 
\frac{\sum_{i=1}^{M_k}(y_{k,i}-\hat{y}_{k,\lambda,i})^2/\sigma_{k,i}^2}
{\sum_{i=1}^{M_k} 1/\sigma_{k,i}^2}\\
\hat{y}_{k,\lambda,i} &=& \sum_{j=1}^N x_{i,j}\hat{\beta}_{k,\lambda,j}
\end{eqnarray}
where $M_k$ is the number of the validation data, $\bm{y}_k$, and
$\sigma_{k,i}$ is the measurement error of the $i$-th element in
$\bm{y}_k$. 

In a very large $\lambda$ regime, the least-square term is large, and
thereby $E(\lambda)$ also becomes large. In a very small $\lambda$
regime, on the other hand, the model can reproduce the noise in the
data (over-fitting), and thereby have a large generalization error,
and eventually lead to a large $E(\lambda)$. Thus, we can find the
minimum value of $E(\lambda)$ at a certain $\lambda$. The best model
can be considered as the simplest model whose $E(\lambda)$ is within
one standard error of the minimal $E(\lambda)$. This is the so-called
``one standard error rule''. Models having $\lambda$ smaller than the
best one are statistically indistinguishable from the over-fitting
situation. In this paper, we use this rule to select $\lambda$, and
set $K=10$. 

Another common variable selection scheme is to use an information
criterion, such as Akaike information criterion (AIC) or Bayesian
information criterion (BIC). We employed the regularization term and
the cross-validation because we expect not only the observation noise
$\bm{e}$ but also the measurement errors in $X$, and we do not have a
good model selection criterion for this situation. The measurement
error of $M_B$ is occasionally quite small, an order of 0.01~mag. On
the other hand, the error of the elements in $X$ can be large. For
example, a ratio between low fluxes can have a large error. 

\subsection{Demonstration of the method}

\begin{figure*}
 \begin{center}
  \includegraphics[width=17cm]{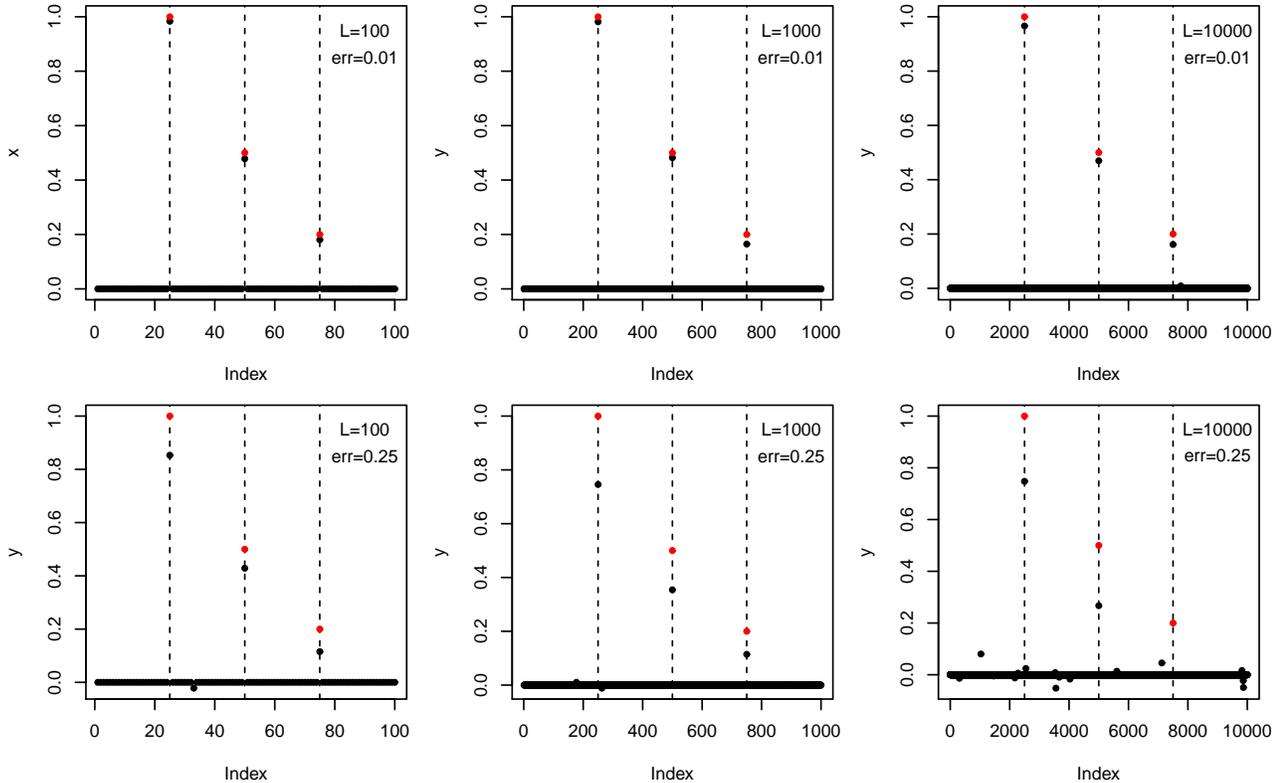} 
 \end{center}
\caption{Simulations of the LASSO-type estimation. The red and black
  points represent the assumed and estimated values,
  respectively. The number of samples is 50. The numbers of
  explanatory variables are $L=10^2$ (left), $10^3$ (middle), and
  $10^4$ (right), as shown in each panel. The upper and lower panels
  depict the cases for small and large errors assumed in $X$.
}\label{fig:demo} 
\end{figure*}

We performed simple simulations of the LASSO-type estimation for
the current problem. The vector, $\bm{\beta}$, was set to be a sparse 
vector, containing only three non-zero values in $L$ elements. We set
three cases: $L=10^2$, $10^3$, and $10^4$. The matrix, $X$, 
was set to be a $N\times L$ matrix whose elements were random values
generated by $\mathcal{N}(0,1)$, a normal distribution with a mean
of $0$ and variance of $1$. We set $N=50$ in all cases. Then, we
calculated the data vector, $X\bm{\beta}$, and added noise, $\bm{y} =
X\bm{\beta}+\bm{e}$,
$\bm{e}\sim\mathcal{N}(\bm{0},0.01\sigma^2_{\bm{y}}I_N)$, where
$\sigma_{\bm{y}}$ represents the standard deviation of observation
noise. Here, we assumed a small error in $\bm{y}$ because $M_B$ is
occasionally determined with such a high precision. We also added
noise in the elements of $X$, $\tilde{x}_{ij}=x_{ij}+\epsilon_{ij}$,
$\epsilon_{ij}\sim\mathcal{N}(0,\sigma_X^2)$, and generated 
$\tilde{X}$. We assumed small and large errors in $X$, that is,
$\sigma_{X}=0.01$ and $0.25$. We estimated $\bm{\beta}$ from $\bm{y}$
and $\tilde{X}$ using the $\ell 1$-norm minimization. The best model
and its $\lambda$ were determined by cross-validation. The results are
shown in figure~\ref{fig:demo}. In the case of the small $\sigma_{X}$,
the assumed $\bm{\beta}$, indicated by the red points, are
successfully reconstructed in all $L$ cases, albeit with a
  $3$--$20$\,\% systematic bias. In the case of the large 
$\sigma_{X}$, all non-zero elements in $\bm{\beta}$ are detected in
the cases of $L=10^2$ and $10^3$, while their coefficients are
significantly underestimated and weak false signals are also seen. In
the case of $L=10^4$, the assumed weak signal is lost in the
reconstruction, and false signals have large coefficients. 

This experiment demonstrated two important points about the proposed
method. First, it can reconstruct the original vector even with the
case of $N<L$. Second, even with this method, we cannot avoid
detecting false signals which are coincidentally fit the data in the
case of a large $L$. The latter point could have a significant
implication for using the arbitrary flux ratios in the current
problem. The number of the flux ratios is more than 17000, while
the number of samples is $\lesssim 100$. Hence, we should reduce the
number of columns in $X$ in order to avoid detecting the false
signals. In this paper, as described in the next subsection, we use two
kinds of spectra normalized by the continuum level and by the total
flux. 

LASSO tends to underestimate the coefficients if the
measurement error of the target variable is not negligible, as can be
seen in figure~\ref{fig:demo}. Hence, it should be used to select the
best set of variables. Then, the model of $M_B$ can be obtained by a
refit to the data with the selected variables. In this paper, we focus
on variable selection.

\subsection{Sample and variables}

We used the data from SuperNova DataBase provided by Berkeley
Supernova Ia
program\footnote{$\langle$http://hercules.berkeley.edu/database/index\_public.html$\rangle$}. 
Our sample selection was based on the criteria in \citet{bsnip3}: The
redshift of the sample ranged from 0.01 to 0.1. We used the spectral
data from 3500 to 8500 \AA. The rest-frame days relative to the
maximum is ranged from $-5$ to $+5\;{\rm d}$. We used the spectrum
having the smallest value of the rest-frame days relative to the
maximum for each object in the case that multiple spectra were
available. We only used samples having the color parameter, $c$, less
than 0.5. We found two Type~Iax objects, SN~2003gq and 2005hk in the
sample, and excluded them (\cite{fol13Iax}). As a result, we found 78
objects in the database. The available data contains, for example, the
redshift, $z$, light-curve width, $x_1$, color, $c$, apparent
magnitude, $m_B$, and spectra. As mentioned in section~1, it is
believed that $x_1$ and $c$ are important explanatory variables for
$M_B$. We calculated $M_B$ from $m_B$ and $z$ by adopting the standard
$\Lambda$ cold dark matter cosmology with $\Omega_m=0.27$,
$\Omega_\Lambda=0.73$, and $w=-1$. 

The calibration of the spectral data was performed in the standard
manner: The flux was corrected for the reddening in our galaxy using
$E(B-V)$. We used the $E(B-V)$ values obtained from the
supernova database, which refers to \citet{sch98dust} and
\citet{pee10dust}. The red-shift correction was performed on the
wavelength. Then, the spectra were divided into 134~bins which were
equally spaced in the logarithmic velocity scale between 3500 and
8500~\AA, as in \citet{bsnip3}. We calculated the arbitrary flux
ratios using the binned spectra. The number of the ratios is then
$134\times 133=17822$. 

Including arbitrary flux ratios may provide an exhaustive search 
for an appropriate set of explanatory variables of $M_B$. However, the
number of candidate variables is so large that false signals can be
detected, as demonstrated in the last subsection. Hence, we need to
consider other sets of candidate variables which are related to the
flux ratios, but have much smaller dimension. In this paper, we use
two kinds of normalized spectra.

\begin{figure}
 \begin{center}
  \includegraphics[width=8.5cm]{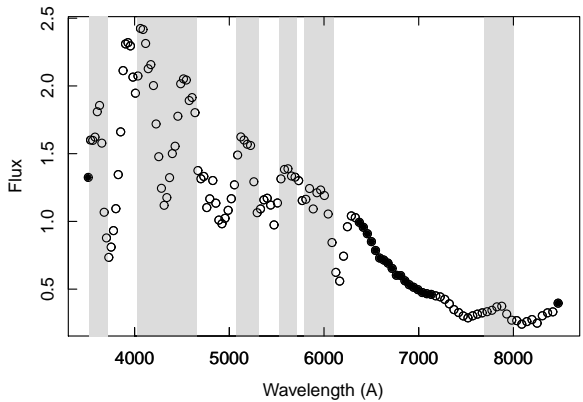} 
 \end{center}
\caption{Mask for calculating the continuum level. The spectrum of
  SN~2006et is also plotted as a reference. For detail, see the
  text.}\label{fig:cont} 
\end{figure}

\begin{figure}
 \begin{center}
  \includegraphics[width=8.5cm]{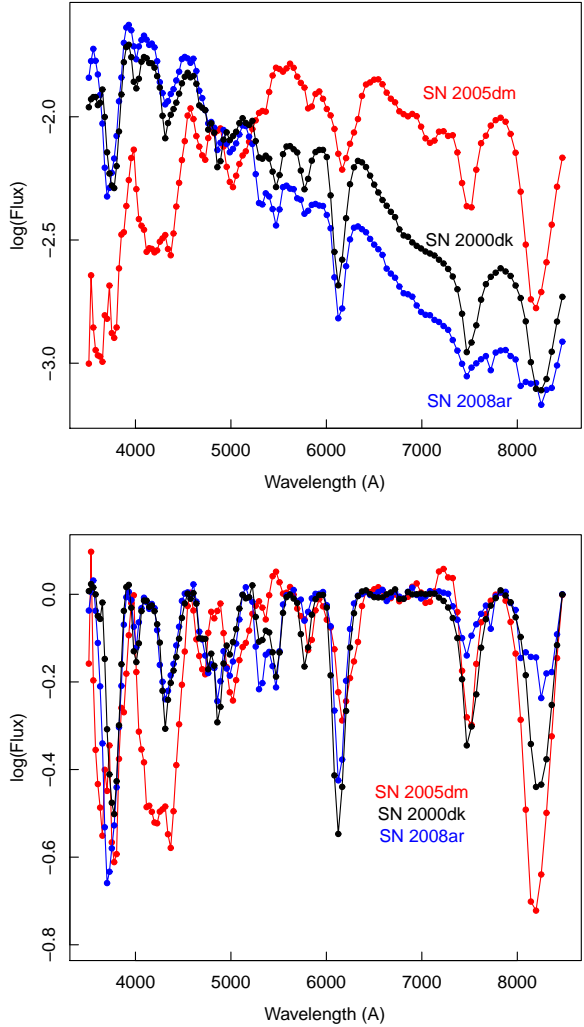} 
 \end{center}
\caption{Examples of the spectra in our sample. Upper panel: the
  spectra normalized by the total flux between 3500--8500\AA. Lower
  panel: the spectra normalized by the continuum. The spectra of three
  examples, SN~2005dm, SN~2000dk, and SN~2008ar, are shown in the both
  panels.}\label{fig:sample}
\end{figure}

First, the variables of the most interest are the flux ratios of the
line areas to the continuum level. Indeed, most of previously proposed 
ratios are such variables: $\mathcal{R}(6420/4430)=$
Fe\,\textsc{ii}/continuum (\cite{bai09frat}), 
$\mathcal{R}(6630/4400)=$ Fe\,\textsc{ii}/continuum, 
$\mathcal{R}(6420/5290)=$ continuum/S\,\textsc{ii}, and 
$\mathcal{R}(4610/4260)=$ continuum/Fe\,\textsc{ii} 
(\cite{blo11frat}). They can be substituted by the spectra normalized
by the continuum level. The continuum level was approximated
by a cubic smoothing spline fitted to masked spectra. The mask is
depicted in figure~\ref{fig:cont} with the binned spectra of a typical
sample, SN~2006et. The data points indicated by the filled circles
were used to calculate the continuum curve. In addition, the points
with the maximum flux in each shaded area were also used. The several
examples of the continuum-normalized spectra are shown in the lower
panel of figure~\ref{fig:sample}. We call the set of the
continuum-normalized spectra as $\bm{f}_{\rm cnt}$. 

Second, the local colors in the continuum which may have independent
information of the broadband colors are also variables of
interest. They can be substituted by the spectra normalized by the
total flux between 3500 and 8500~\AA. We call the set of this total
flux normalized spectra as $\bm{f}_{\rm tot}$. The intrinsic color can
be bluer than the observed one because of the interstellar reddening
effect in the host galaxy. In previous studies, the color correction
for this effect has been performed by assuming that all SNe~Ia have
the same intrinsic color. We also performed this correction using the
SNe~Ia color-law for the SALT2 data (\cite{guy07salt2}). The
color-corrected spectra is, then normalized by the total flux, and
named $\bm{f}^c_{\rm tot}$.  

We include those two kinds of normalized spectra, 
$(\bm{f}_{\rm cnt}, \bm{f}_{\rm tot})$ or $(\bm{f}_{\rm cnt},
\bm{f}^c_{\rm tot})$ as the candidates, instead of the arbitrary flux
ratios. In addition, we use the flux in the logarithmic scale in 
order to include the information of arbitrary flux ratios. We can
identify a good flux-ratio parameter by searching for the two fluxes
having the similar coefficients with the opposite sign: $c \cdot
\log(f_1/f_2)=c \cdot \log(f_1) - c \cdot \log(f_2)$. 
Figure~\ref{fig:sample} shows examples of the spectra that are
normalized by the total flux ($\bm{f}_{\rm tot}$, the upper panel) and
by the continuum ($\bm{f}_{\rm cnt}$, the lower panel). 

As well as $x_1$, $c$, $\bm{f}_{\rm cnt}$, $\bm{f}_{\rm tot}$, and
$\bm{f}^c_{\rm tot}$, we include previously proposed flux-ratios, 
$\bm{\mathcal{R}}$ into the model as candidate explanatory variables
for $M_B$. We consider six flux-ratios proposed in \citet{bai09frat},
\cite{blo11frat}, and \citet{bsnip3}, that is, 
$\bm{\mathcal{R}}=\{\mathcal{R}(3780/4580),\;
\mathcal{R}(4610/4260),\; \mathcal{R}(5690/5360),$   
$\mathcal{R}(6420/4430),\; \mathcal{R}(6420/5290),\;
\mathcal{R}(6630/4400)\}$. The flux ratios which are calculated from
the color-corrected spectra, $\bm{f}^c_{\rm tot}$, are called as
$\bm{\mathcal{R}}^c$. 

\citet{bsnip3} presents tables of measured values of the lines:
Ca\,\textsc{ii} H\&K and near-infrared triplet, Si\,\textsc{ii}
4000, 5972, and 6355~\AA, Mg\,\textsc{ii}, Fe\,\textsc{ii},
S\,\textsc{ii} ``W,'' and O\,\textsc{i} triplet. We can use pEW, Delta
pEW (i.e., the measured pEW subtracted by the template evolution),
velocity ($v$), line depth ($a$), and FWHM for the explanatory 
variables. We note that those line variables are incomplete for our
sample. Hence, the number of samples reduces when those line
variables are used as the candidate variables, and we used them
element by element. We represent a set of the line values as
$\bm{\mathcal{L}}$. For example, $\bm{\mathcal{L}}_
{\rm Si\,\textsc{ii} 4000}$ means those variables of Si\,\textsc{ii}
4000\AA. 

For the optimization of the model to the data, we used the
\texttt{glmnet} package for \texttt{R}.
\footnote{$\langle$http://www.r-project.org/$\rangle$} The selection
of $\lambda$ was performed using the function for the cross-validation,
\texttt{cv.glmnet}, adopting the one-standard error rule. The
cross-validation is based on random sub-sampling and the selected
variables might be influenced by it. We performed $10^4$ experiments
for each model, and calculated the selection probability, $p$, of each
variable. In this paper, we discuss selected variables only with
$p>0.3$. Each column in $X$ was normalized to have zero mean and unit
variance, by a linear scaling,
$x^\prime_{ij}=(x_{ij}-\bar{x}_j)/\sigma_j$, where $\bar{x}_j$ and
$\sigma_j$ are the mean and standard deviation of the $j$-th
column. We need this normalization to compare the coefficients,
$\bm{\beta}$, of variables having different units. \textbf{The list of
objects and explanatory variables used in this paper is available as
an online supplement material.}

\section{Results}

\begin{table*}
  \caption{Models and Results}\label{tab:model}
  \begin{center}
    \begin{tabular}{lrrrrr}
      \hline
      Model & Target variable & Explanatory variables & Non-zero
      elements & coefficients & $p$ \\
      & $\bm{y}$ $(N)$& $X$ $(L)$& &$\bm{\beta}$ &\\ 
      \hline
      1 & $M_B$ $(78)$& $x_1,c,\bm{f}_{\rm tot},\bm{f}_{\rm
      cnt},\bm{\mathcal{R}}$ $(276)$ &
              $c$              &$0.376$ & 1.00\\
        & & & $f_{\rm tot}(6373)$&$0.100$ & 1.00\\
        & & & $x_1$            &$-0.050$& 0.98\\
        & & & $f_{\rm cnt}(6084)$&$-0.034$& 0.98\\
        & & & $f_{\rm cnt}(6289)$&$-0.045$& 0.95\\
        & & & $f_{\rm cnt}(6631)$&$-0.061$& 0.80\\
        & & & $\mathcal{R}(3780/4580)$&$-0.050$& 0.74\\
        & & & $f_{\rm tot}(3752)$&$0.063$& 0.73\\ \hline
      2 & $M_B-\beta_1 c$ $(78)$ & $x_1,\bm{f}_{\rm tot},\bm{f}_{\rm
        cnt},\bm{\mathcal{R}}$ $(275)$ & 
              $x_1$            &$-0.020$& 0.99\\ \hline
      3 & $M_B-\beta_1 c$ $(78)$ & $x_1,\bm{f}^c_{\rm tot},\bm{f}_{\rm
        cnt},\bm{\mathcal{R}}^c$ $(275)$ &
              $x_1$            &$-0.014$& 0.85\\ \hline
      4a & $x_1$ $(76)$& $c,\bm{f}^c_{\rm tot}, \bm{f}_{\rm cnt},
      \bm{\mathcal{R}}^c,\bm{\mathcal{L}}_{\rm Si\,II\,4000}$ (280)&
              ${\rm DpEW}_{\rm Si\,II\,4000}$&$-0.455$& 1.00\\
        & & & $f_{\rm cnt}(5770)$&$0.518$ & 1.00\\
        & & & $f_{\rm cnt}(3982)$&$-0.262$& 1.00\\
        & & & $f_{\rm cnt}(7038)$&$-0.485$& 0.96\\
        & & & $f^c_{\rm tot}(4988)$&$-0.238$& 0.77\\
        & & & $f_{\rm cnt}(6084)$&$0.281$ & 0.62\\ \hline
      4b & $x_1$ $(74)$ & $c,\bm{f}^c_{\rm tot}, \bm{f}_{\rm cnt},
      \bm{\mathcal{R}}^c,\bm{\mathcal{L}}_{\rm S\,II ``W''}$ (280)&
              $f_{\rm cnt}(5770)$& $1.034$& 1.00\\
        & & & $f_{\rm cnt}(6084)$& $0.440$& 1.00\\
        & & & $f^c_{\rm tot}(6458)$& $0.300$& 1.00\\
        & & & $f_{\rm cnt}(3982)$& $0.041$& 1.00\\
        & & & $f_{\rm cnt}(7179)$& $0.289$& 0.99\\
        & & & $f_{\rm cnt}(6458)$&$-0.236$& 0.94\\
        & & & $f_{\rm cnt}(6331)$& $0.612$& 0.92\\ \hline
      5 & $M_B-(\beta_1 c + \beta_2 x_1)$ $(78)$ & $\bm{f}^c_{\rm
        tot},\bm{f}_{\rm cnt},\bm{\mathcal{R}}^c$ $(273)$ & 
              --- & --- & ---\\
      \hline
    \end{tabular}
  \end{center}
\end{table*}

\begin{figure}
 \begin{center}
  \includegraphics[width=8.5cm]{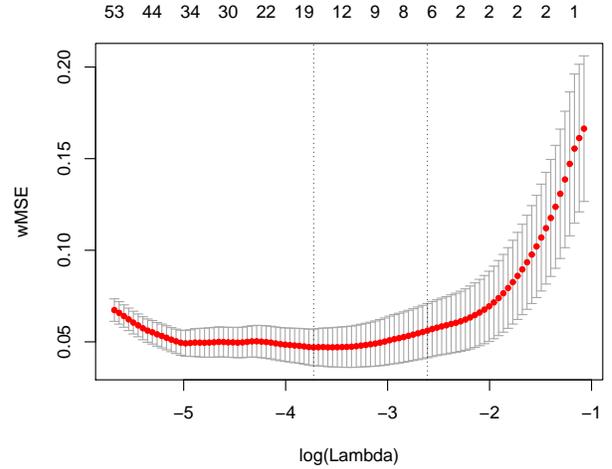} 
 \end{center}
\caption{Cross-validation curve for Model~1. The lower and upper
  horizontal axes denote $\lambda$ and the number of non-zero
  elements, respectively. The vertical axis denotes
  wMSE. The left dotted line indicates $\lambda$ having the minimal
  wMSE. The right dotted line indicates the best model under the
  one-standard error rule.}\label{fig:cv_model1} 
\end{figure}

First, we choose the light curve width ($x_1$), color ($c$), spectra
normalized by the total flux ($\bm{f}_{\rm tot}$), those by the
continuum ($\bm{f}_{\rm cnt}$), and previously proposed flux-ratios
($\bm{\mathcal{R}}$) as the candidate explanatory variables, and $M_B$
as the target variable. We call this complete model as Model~1. It can
be rewritten as: 
\begin{eqnarray}
 M_B &=& M_{B,0} + \beta_1 c + \beta_2 x_1 \nonumber \\
 &+& \beta_3 f_{\rm tot}(3512) + \beta_4 f_{\rm tot}(3534) + \cdots + \beta_{136} f_{\rm tot}(8472) \nonumber\\
 &+& \beta_{137} f_{\rm cnt}(3512) + \beta_{138} f_{\rm cnt}(3534) + \cdots + \beta_{270} f_{\rm cnt}(8472) \nonumber \\
 &+& \beta_{271} \mathcal{R}(3780/4580) + \beta_{272} \mathcal{R}(4610/4260) \nonumber \\ 
 &+& \beta_{273} \mathcal{R}(5690/5360) + \beta_{274} \mathcal{R}(6420/4430) \nonumber \\
 &+& \beta_{275} \mathcal{R}(6420/5290) + \beta_{276} \mathcal{R}(6630/4400) +e.
 \end{eqnarray}
Using LASSO-type method for 78 samples of $M_B$, we choose the
appropriate set of explanatory variables from 276 candidates and
estimate coefficients vector $\bm{\beta}$. The tuning parameter,
$\lambda$, is determined by cross-validation. 
Figure~\ref{fig:cv_model1} shows the cross-validation curve for
Model~1. In this figure, we can confirm that wMSE take the minimum
value in a given range of $\lambda$, and the best model is properly
determined by the one-standard-error rule.

\begin{figure}
 \begin{center}
  \includegraphics[width=8.5cm]{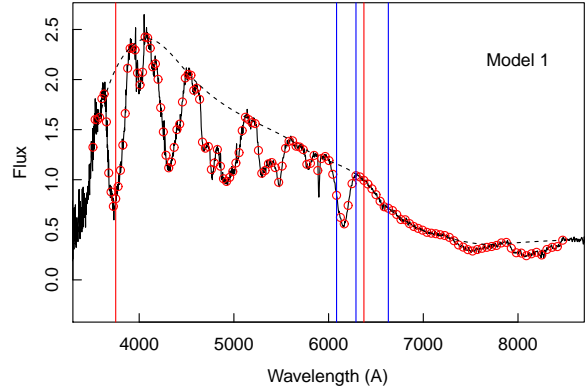} 
 \end{center}
\caption{Non-zero elements of spectral data in Model~1. The red and
  blue lines indicate the non-zero elements in the
  total-flux-normalized and continuum-normalized spectra,
  respectively. The spectrum of SN~2006et is also plotted as a
  reference. The solid and dashed lines and red points are unbinned
  spectra, estimated continuum level, and binned spectra,
  respectively.}\label{fig:model1} 
\end{figure}

Table~\ref{tab:model} lists all models and results presented in this
paper. From Model~1, the classical variables, that is, $c$ and $x_1$
are selected. $f_{\rm tot}(6373)$ is also selected, having a
coefficient even larger than that of $x_1$ in the absolute values. 
Figure~\ref{fig:model1} indicates the non-zero elements of  
$\bm{f}_{\rm tot}$ and $\bm{f}_{\rm cnt}$. As can be seen in this
figure, $f_{\rm tot}(6373)$, indicated by the red vertical line, lies
in the continuum area. Hence, it may be related to the local color
which could have specific information against the broad-band color,
$c$. As can be seen in figure~4, some fluxes in line regions
are also selected: $f_{\rm cnt}(6084)$ and $f_{\rm cnt}(6289)$ are
probably related to the continuum-normalized depths of
Si\,\textsc{ii}(6355). In addition, $\mathcal{R}(3780/4580)$ and
$f_{\rm tot}(3752)$ are probably relate to Ca\,\textsc{ii} H\&K. 
$f_{\rm cnt}(6631)$ corresponds to the continuum flux of the
continuum-normalized spectra, which suggests a false signal.

We confirmed that the result was consistent even when we included the
line variables, $\bm{\mathcal{L}}$, and all the line variables have zero
coefficients in any elements. The lack of the dependency on
$\bm{\mathcal{L}}$ is common in other subsequent models, except for
Model~4 (see below). Hence, we present the models only without
$\bm{\mathcal{L}}$ in this paper. 

In general, when some explanatory variables are correlated,
LASSO could select a few of them. In the present case, the variables
are measurements having non-negligible errors. Among correlated
variables, a variable having a smaller error results in a smaller
generalization error of $M_B$. Hence, in the case of large $\lambda$,
the variable having the smallest error is first selected. In the case
of smaller $\lambda$, the other correlated variables are selected. It
is possible that a high correlation between $c$ and $f_{\rm tot}(6373)$ 
would cause the non-zero coefficient of $f_{\rm tot}(6373)$ in
Model~1. If this is the case, it is unclear that $f_{\rm tot}(6373)$
is a significant variable that has independent information of $c$.
We performed a regression analysis with $M_B=\beta_1 c + M_{B,0}$, and
corrected for the effect of $c$ in $M_B$ by using $M_B-\beta_1 c$ as
the target. We call this complete model as Model~2. The number of
samples is the same as that of Model~1, 78, while the number of
candidate explanatory variables is 275, one smaller than in Model~1
because $c$ is omitted. As can be seen in table~\ref{tab:model}, $x_1$
is the only variable having a non-zero coefficient. A similar result
was obtained for Model~3, where we used the color-corrected spectral
data, $\bm{f}^c_{\rm tot}$ and $\mathcal{R}^c$ instead of $\bm{f}_{\rm
  tot}$ and $\mathcal{R}$. Hence, the lack of $f_{\rm tot}(6373)$ is
independent of the color correction. These results suggest that the
high correlation between $c$ and $f_{\rm tot}(6373)$ causes the
apparently high coefficient of $f_{\rm tot}(6373)$ in Model~1. 

\begin{figure}
 \begin{center}
  \includegraphics[width=8.5cm]{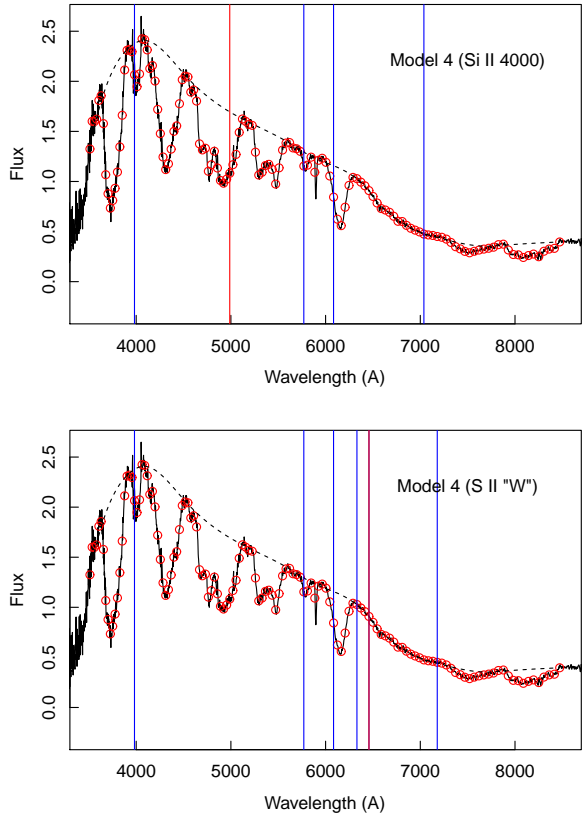} 
 \end{center}
\caption{Same as figure~\ref{fig:model1}, but for Models~4a and
  4b.}\label{fig:model_x1} 
\end{figure}

As well as $f_{\rm tot}(6373)$, Model~1 indicates possible dependency
of $M_B$ on the variables related to the line areas, that is, 
$f_{\rm cnt}(6084)$ and $f_{\rm cnt}(6289)$. It has been reported that
$x_1$ depends on the line strength, for example, the EW of
Si\,\textsc{ii}~4000 (e.g. \cite{hac06snsp,ars08x1}). It is possible
that the line dependency in Model~1 may be due to a high correlation
between $x_1$ and the line strength of Si\,\textsc{ii}.  
For examining this possibility, we considered Model~4 in which the
target is $x_1$. Model~4a includes pEW, DpEW, $v$, $a$, FWHM of
Si\,\textsc{ii} 4000, as well as $c$, $\bm{f}^c_{\rm tot}$, $\bm{f}_{\rm
  cnt}$, and $\bm{\mathcal{R}}^c$ as the candidate explanatory
variables of $x_1$. This is the unique case where the coefficients of
$\bm{\mathcal{L}}$ have non-zero values in the analysis presented in
this paper. Model~4b is for S\,\textsc{ii} ``W'', as a typical case of
the other lines. The results are shown in table~\ref{tab:model} and
figure~\ref{fig:model_x1}. In Model~4a, DpEW of Si\,\textsc{ii}~4000
has a non-zero coefficient. The importance of this line is also
confirmed by the selection of $f_{\rm cnt}(3982)$ both in Models~4a 
and b. In addition to Si\,\textsc{ii}~4000, $f_{\rm cnt}(5770)$ and
$f_{\rm cnt}(6084)$ have non-zero coefficients in both models,
corresponding to Si\,\textsc{ii} 5972 and 6355. There are several
other non-zero elements in Model~4a, although they are not confirmed
in Model~4b. The dependence of $x_1$ on Si\,\textsc{ii} supports the
previous studies about $x_1$.  

Finally, we employed Model~5, in which the target is $M_B$ corrected
for $c$ and $x_1$, that is, $M_B-(\beta_1 c + \beta_2 x_1)$, where
$\beta_1$ and $\beta_2$ are determined by a regression analysis. The
candidate explanatory variables of Model~5 are $\bm{f}^c_{\rm tot}$,
$\bm{f}_{\rm cnt}$, and $\bm{\mathcal{R}}^c$. However, any of them is
not selected. The result suggests that the high correlation between
$x_1$ and the Si\,\textsc{ii} line strength results in the apparent
dependency of $M_B$ on the line depths in Model~1. Hence, the
best set of explanatory variables is $(c,x_1)$ in our analysis. We
re-fit the data with these variables, and obtained the following
model:
\begin{eqnarray}
M_B = -19.26(\pm 0.03) + 2.75 (\pm 0.17) c - 0.10 (\pm 0.02) x_1
\end{eqnarray}
Note that these values are calculated not from normalized values of
the variables as in table~1, but from raw values.

\section{Discussion and Conclusion}

Our analysis confirms the classical understanding of SNe~Ia, that is,
i)~the light-curve width ($x_1$) and color ($c$) are the important
explanatory variables of the absolute magnitude at maximum ($M_B$)
(\cite{phi93law}), and ii)~the light-curve width correlates with the
strength (EW or depth) of Si\,\textsc{ii}
(e.g. \cite{hac06snsp,ars08x1}). Furthermore, our variable selection
approach using the LASSO-type estimation does not support to add any
other variables, such as the normalized spectra ($\bm{f}_{\rm tot}$,
$\bm{f}^c_{\rm tot}$, $\bm{f}_{\rm cnt}$), previously proposed flux
ratios ($\bm{\mathcal{R}}$), and line measurements
($\bm{\mathcal{L}}$), in order to have a better generalization error
of $M_B$. We confirmed that the above conclusion is robust to small
changes in our analysis: using the flux in logarithmic or linear scale,
excluding or including two Type~Iax objects, and normalizing each
column in $X$ or not. Our analysis implies that over-fitting
can cause partly inconsistent results seen in previous studies which used
the arbitrary flux ratios (\cite{bai09frat,blo11frat,bsnip3}). 

\begin{figure}
 \begin{center}
  \includegraphics[width=8.5cm]{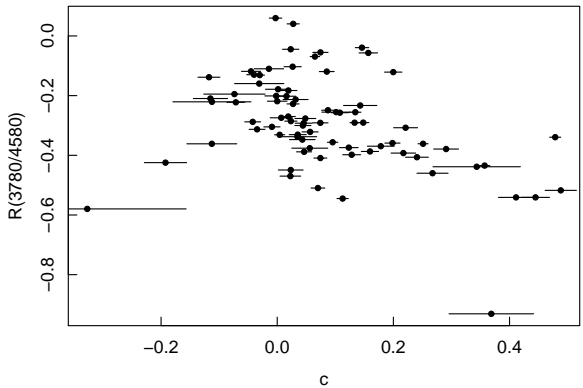} 
 \end{center}
\caption{$\mathcal{R}^c(3780/4580)$ of
our sample against $c$}\label{fig:frat-c} 
\end{figure}

Our conclusion is inconsistent with that reported by \citet{bsnip3}
although the both samples are obtained from the Berkeley supernova
database with the common data selection. The model selection method
is also common: Following \citet{blo11frat}, \citet{bsnip3} performed
10-fold cross-validation, and calculated the mean and standard error
of 10 weighted root-mean squares (wRMS) of residuals. They measured
the significance of the improvement of the model with the mean wRMS
and its standard error. They found that the model with $c$, $x_1$, and
$\mathcal{R}^c(3780/4580)$ improves the prediction error by a level of 
$1.7\sigma$ compared with the classical one with $c$ and $x_1$. This
flux ratio is also detected as an explanatory variable in our Model~1,
while it is not in the other models of $M_B$ (see table~1). The
wavelength of 3780\,\AA~corresponds to the mid-point of
Ca\,\textsc{ii} H\&K, and 4580\,\AA~to the border between the 
Mg\,\textsc{ii} and Fe\,\textsc{ii} complexes. Figure~\ref{fig:frat-c}
shows $\mathcal{R}^c(3780/4580)$ of our sample against $c$. Those two
variables exhibit a weak anticorrelation, as can be seen in this
figure. Our result that $\mathcal{R}^c(3780/4580)$ is detected in
Model~1 and not in Models~2 and 3 can be explained by this
anticorrelation. 

In principle, the spectral data, that is, the values of the flux
density and their ratios, could be important explanatory variables of
$M_B$. The interstellar extinction is definitely the most important
variable. As well as the color parameter, $c$, the continuum flux of
the total-flux normalized spectra, that is, $\bm{f}_{\rm tot}$ could
be an indicator of the extinction. Indeed, $f_{\rm tot}(6373)$ has
a relatively large coefficient in Model~1. Our analysis suggests that
$c$ is a better variable rather than $f_{\rm tot}(6373)$ and other
normalized fluxes. This is probably because of the uncertainty of
measurements and observation epochs. In general, the flux calibration
of spectra has larger errors than the differential
photometry. Moreover, the observation epochs of spectra are different 
from one object to another in our sample. The color parameter, $c$, is
based on differential photometry, and corrected to the color at
maximum. A similar situation is also expected in the light-curve
width, $x_1$. \citet{maz01decl} propose that $x_1$, or decline rate,
so-called, $\Delta m_{15}$ is a function of the amount of $^{56}{\rm
  Ni}$ produced in SNe. The absorption line variables are also
possible indicators of the amount of synthesized elements in
SNe. Indeed, the correlation between $x_1$ and the strength of the
Si\,\textsc{ii} lines was confirmed in Model~4. Our method selected
not the Si\,\textsc{ii} strength, but $x_1$ probably because of the
small measurement error in $x_1$ for the amount of synthesized
elements. In our analysis, the best set of the explanatory variables
is $c$ and $x_1$, while it is trivial that our result does not imply a
physical causal relationship between $M_B$ and those two variables. 

It is possible that, in future, the increasing number of samples
revises the model having a better generalization error by finding
additional or alternative explanatory variables compared with the
model in this paper. It may also be meaningful to add variables which 
were not used in this paper, such as those about host galaxies of
SNe~Ia (\cite{sul10host,pan15host}). In any of these cases, our
proposed method offers a framework for finding an appropriate set of
explanatory variables of $M_B$ even in the case that the number of
samples is smaller than the number of variables. A possible
extension of the model may be to include the measurement errors of
explanatory variables. As can be seen in equation~(2), our method does
not include the errors, while errors are expected to be large in
several variables, for example, flux ratios of low fluxes. 
\cite{bai09frat}, proposing the flux ratio, $\mathcal{R}(642\,{\rm
  nm}/443\,{\rm nm})$, as a good explanatory variable of $M_B$,
claimed that the spectral slope need to be calibrated with very small
errors for their model with the flux ratio. The instrument that they
used was SNIFS (SuperNova Integral Field Spectrograph), which was
developed to perform flux-calibration with a high accuracy
(\cite{ald02snif}). On the other hand, the instruments which were used
in \citet{blo11frat} and \citet{bsnip3} were standard slit
spectrographs. It is possible that the lack of detection of
$\mathcal{R}(642\,{\rm nm}/443\,{\rm nm})$ in our analysis is due to
large errors of the flux ratios in our data sample from
\citet{bsnip3}. A better model including the errors might be provided
by a Bayesian approach in which the error is included into the model
as prior probability distributions.

We would like to thank Drs. J. M. Silverman and A. V. Filippenko for
providing the Berkeley supernova database. We also appreciate comments
and suggestions from anonymous referee.
This work was supported by JSPS KAKENHI Grant Number 25120007,
25120008, and 26800100. The work by K.M. is partly supported by WPI
Initiative, MEXT, Japan.

\end{document}